\documentclass{aa}  
\usepackage{graphicx}
\usepackage{txfonts}
\usepackage[colorlinks=true,
    linkcolor=blue,
    citecolor=blue,
    filecolor=magenta,      
    urlcolor=cyan]{hyperref}
\newcommand{\msun}{ M$_\odot$ }

\newcommand{\lsim}{\mathrel{\rlap{\lower 3pt \hbox{$\sim$}} \raise 2.0pt \hbox{$<$}}}
\newcommand{\gsim}{\mathrel{\rlap{\lower 3pt \hbox{$\sim$}} \raise 2.0pt \hbox{$>$}}}

\begin{document} 

   \title{A fast test for the identification and confirmation of massive black hole binaries}
   \titlerunning{A fast test for MBHB identification}
   \authorrunning{M. Dotti et al.}

   \author{Massimo Dotti
          \inst{1,}
          \inst{2,} \inst{3}\fnmsep\thanks{massimo.dotti@unimib.it}
          \and
          Fabio Rigamonti
          \inst{4,}
          \inst{2,}
          \inst{3}
          \and
          Stefano Rinaldi
          \inst{5,6}
          \and
          Walter Del Pozzo
          \inst{5,6}
          \and 
          Roberto Decarli
          \inst{7}
          \and
          Riccardo Buscicchio          \inst{1,2}
    }

   \institute{
            Universit\`a degli Studi di Milano-Bicocca, Piazza della Scienza 3, 20126 Milano, Italy
        \and
            INFN, Sezione di Milano-Bicocca, Piazza della Scienza 3, I-20126 Milano, Italy
        \and
            INAF - Osservatorio Astronomico di Brera, via Brera 20, I-20121 Milano, Italy
        \and
            DiSAT, Universit\`a degli Studi dell'Insubria, via Valleggio 11, I-22100 Como, Italy
        \and
            Dipartimento di Fisica ``E. Fermi'', Università di Pisa, I-56127 Pisa, Italy
        \and
            Istituto Nazionale di Fisica Nucleare, Sezione di Pisa, I-56127 Pisa, Italy
        \and
            INAF – Osservatorio di Astrofisica e Scienza dello Spazio di Bologna, Via Gobetti 93/3, I-40129 Bologna, Italy
    }

   \date{Received XXX; accepted YYY}

 
\abstract{
We present a new observational test to identify massive black hole binaries in large multi-epoch spectroscopical catalogues and to confirm already proposed binary candidates. The test is tailored for binaries with  large
enough separations to allow each black hole to retain its own broad line region (BLR). Within this limit, the fast  variability of active galactic nuclei (AGN) typically observed over months cannot be associated to the much longer binary period and is assumed (as for the case of single black holes) to be the consequence of the evolution of the innermost regions of the two accretion discs. A simple analysis of the cross-correlation between different parts of individual broad emission lines can therefore be used to identify the presence of two massive black holes whose continua vary independently of each other. Our analysis indicates that, to be less affected by the noise in the spectra, the broad lines should be divided into two  parts of almost  equal flux. This ensures that, in the single massive black hole scenario, the cross-correlation will always be strong. With monitoring campaigns similar to those performed for reverberation mapping studies, inversely,  a binary can show any value of the cross-correlation and can therefore be distinguished from a standard  AGN.
  This new test can be performed over timescales  that are orders of magnitude shorter than the alternative tests already discussed in the literature, and can be a powerful complement to the massive black hole binary search strategies already in place.
}


   \maketitle
%
\section{Introduction}

Gravitationally bound massive black hole binaries (MBHBs) are the  outcome of galaxy mergers, and were originally predicted by \cite{BBR80}. They are among the loudest sources of gravitational waves (GWs) detectable with current pulsar timing array (PTA) campaigns \citep{pta} and by future space-based gravitational wave interferometers \citep[e.g. LISA][]{lisa1,lisa2}. 
The GW background generated by MBHBs, and the rate of detectable binary coalescence, are still quite uncertain. This is partly a consequence of the uncertainties on the efficiency of MBH pairing during the early stages of galaxy mergers \citep{fiacconi13,delvalle15,tamburello17,souzalima17,bortolas20,bortolas22}. The electromagnetic (EM) identification of a sample of already bound MBH binaries would therefore greatly reduce the uncertainties on the signals observable with both PTA and LISA.

To date, unequivocal observational confirmation of an MBHB has not yet been found. The only spatially resolved MBHB candidate is 0402+379 \citep{Rodriguez09, BurkeSpolaor11}, which has two flat-spectrum radio cores with a projected separation of $\approx 7$ pc. 

At smaller, spatially unresolved scales, MBHBs have been consistently searched for either through peculiar spectral features \citep{Tsalmantza11, Eracleous12, Ju13, Shen13, Wang17} or through photometric variability  (\citealt{Valtonen08, Ackermann15, Graham15, Li2016, Charisi16, Sandrinelli16,Sandrinelli18, Severgnini18, Li+2019, LiuGez+2019,Chen+2020}).
These two search strategies target MBHBs at different separations. The spectroscopic approach assumes that the broad line region (BLR) of at least one of the binary components does not extend beyond its MBH Roche radius. Therefore, by sharing the same kinematics as the MBH, the broad emission lines (BELs) are shifted in frequency with respect to the host galaxy rest frame and evolve in time over a binary orbital period $\tau_{\rm orb}$. The constraint on the  size of the BLR can be translated into a constraint on the minimum binary separation, which is of the order of $\sim 0.1$ pc for a $10^8$ \msun MBH accreting at one-tenth of its Eddington ratio (see section \ref{sec:analytics} for more details).

At smaller separations, the BLR is either truncated by the time-dependent binary potential \citep{Montuori11} or shared by both MBHs and comoving with the MBHB centre of mass. Theoretical studies predict periodic variability on timescales of $\sim \tau_{\rm orb}$, either associated with the periodic fueling from the circumbinary material \citep[e.g.][]{HMH08} or with the Doppler boosting of the emission for very close binaries \citep{DHS15}. At close separations, such timescales can be as short as $\lsim 1$ yr, allowing observational searches in current and future multi-epoch observations \citep[see][and references therein]{Graham15, Charisi16, LiuGez+2019,Chen+2020}.

Nevertheless, a major problem is that all of the above-mentioned features used to  identify MBHB candidates can have alternative explanations \citep[see][for an overview of the possible alternative interpretations]{dotti23}.  Theoretical prediction of clear and unique observational features associated with MBHBs is therefore needed in order to test the actual binary nature of both small- (selected through their variability) and large-scale  (spectroscopically identified) MBHB candidates. Some tests have recently been proposed for the small-scale MBHB candidates, including the possibility of periodic gravitational lensing from the companion of the active component of the MBHB \citep[for  binaries observed nearly edge-on, see e.g.][]{doraziolense18, davelaar22a, davelaar22b} and periodic evolution of the polarisation fraction and angle \citep{polar22}.

For MBHBs at  larger separations  \citep[e.g.][]{gaskell96, eracleous97}, a straightforward test consists in observing the expected Doppler drift in the BEL profiles according to the motion of the active component (or components, if both MBHs are active) of the binary with a  $\tau_{\rm orb}$ period \citep[e.g.][]{gaskell96, eracleous97}. A conclusive test would require following the system spectroscopically for $\sim \tau_{\rm orb}$. However, at these scales,  $\tau_{\rm orb}$ can be as high as $\gsim 10 - 100$ yr (see section \ref{sec:analytics}), making such a test (dubbed slow periodicity test (SPT) hereafter) challenging and unpractical for some of the candidates \citep[see e.g.][]{Eracleous12, decarli13, runnoe17}.

In this paper, we propose a new test for the same large-separation binary candidates based on the assumption that, at such large separations, the short-timescale 
($\lsim 1$ day) intrinsic accretion variability is unrelated to the binary period, and that, when both MBHs are active, their variability patterns are uncorrelated.
Each BLR reverberates to the varying continuum of its MBH \citep[e.g.][]{BmK}.  As we argue in the following, dividing the BELs into two components (one `red' and one `blue', which together account for the total flux in the BEL) and cross-correlating the time-evolution of their fluxes can provide a means to identify MBHBs.
Indeed, in the binary scenario, each observed BEL would be composed into two different broad lines, each  with a different velocity offset with respect to the galaxy rest frame, and with the two components varying independently. In this scenario, the cross-correlation between the blue and red parts of the BELs can be significantly smaller with respect to the case of an AGN powered by a single MBH, as long as the velocity offset is not negligible (i.e. each component associated to one of the MBHs contributes significantly to one of the blue or red parts of the global BEL). This implies that our test becomes decreasingly effective for larger and larger binary separations, with a cross-correlation between the two different parts of the BEL getting closer and closer to the high values expected for single MBHSs, and failing to detect any true binary for close-to-zero relative line-of-sight velocity between the two MBHs (see section 4).  
However, while other proposed spectroscopical tests are expected to fail when the velocity shift between the two components is equal to their width \citep[e.g.][]{shenloeb10}, in section 4 we demonstrate that our test can still identify binaries at larger separations and smaller velocity shifts, including binaries with periods of up to $\sim 1000$ yr.

Our proposed method can be thought of as a simplified version of the tests proposed by \cite{wang18, songsheng20}, in which a catalogue of two-dimensional transfer functions (TFs) for different binaries is constructed and compared to the full TF of a candidate binary. While our method is clearly less sensitive to the details of the BEL profile variability, it provides fast and quantitative verification of the binary nature of the candidate, even when the quality of the data is suboptimal for constraining the details of the TF.

The clearest advantage of our new test (dubbed fast uncorrelated variability test; FUVT) is that it can be performed over timescales comparable to the typical duration of the reverberation mapping (RM) campaigns \citep[e.g.][]{bentz09a}, which is usually smaller than 1 yr (down to weeks, and orders of magnitude smaller than $\tau_{\rm orb}$). A second advantage is that, while FUVT can be applied to already identified MBHB candidates with shifted asymmetric or double-peaked BELs, it can also be used to identify new binaries with apparently `standard' BELs in large multi-epoch samples of AGNs.

This paper is organised as follows: in \S~\ref{sec:analytics} we describe the timescales required by FUVT and its range of applicability in terms of intrinsic MBHB properties; in \S~\ref{sec:single} we describe how the FUVT is structured, and we present the results of its application to a real sample of type I AGN (sample I hereafter) studied using RM; in \S~\ref{sec:mock} we explain how we constructed a mock catalogue of MBHBs (sample II) starting from sample I, and present the results of the application of FUVT to the new binary sample; we then conclude with \S~\ref{sec:discussion}, summarising the main results of our study, describing the advantages and disadvantages of FUVT with respect to SPT, and discussing how future observations can further strengthen the test results.

\section{Analytical estimates of FUVT range of applicability}\label{sec:analytics}

As mentioned in Sect. 1, both SPT and FUVT assume that both the BLRs are bound to and comoving with  the individual components of the binary. The BLR radius depends on the BEL we are focusing on and on the luminosities of the accreting MBHs. For the broad H$\beta$ line used in the following, the BLR radius is \citep{bentz09b}
\begin{equation}\label{eq:bentz}
    R_{\rm B-H\beta} \approx 34 \, {\rm light \, day} \times \left(\frac{\lambda L_{\lambda, 5100}}{10^{44} {\rm \, erg \, s^{-1}}}\right)^{0.519},
\end{equation}
where $\lambda L_{\lambda, 5100}$ is the monocromatic luminosity of the AGN continuum at 5100 \AA. Assuming that the bolometric luminosity is $L_{\rm bol}\approx 9 \, \lambda L_{\lambda, 5100}$ \citep{kaspi00}, we can express eq.~\ref{eq:bentz} as a function of the individual MBH mass $M$ and its Eddington ratio $f_{\rm Edd} = L_{\rm bol}/L_{\rm Edd}$:
\begin{equation}\label{eq:bentz2}
    R_{\rm B-H\beta} \approx 11\, {\rm light \, day} \times \left(f_{\rm Edd} \, \frac{M}{10^6\,{\rm M}_\odot} \right)^{0.519}.
\end{equation}
It is   required that this radius be smaller than the Roche lobe radius of each individual MBH \citep{Montuori11}. 
For the test to work, the fluxes from the two accretion discs (and therefore the BLR radii) have to be comparable. For this reason, the minimum separation between the two MBHs is set by the Roche lobe of the secondary MBH, which for circular binaries is  \citep{eggleton83} 
\begin{equation}\label{eq:RL}
    R_{\rm RL,2}\approx 0.49 \, a \, \frac{ q^{2/3}}{0.6\, q^{2/3}+\ln (1+q^{1/3})},
\end{equation}
where $a$ is the separation between the two MBHs and $q=M_2/M_1$ is the ratio between the secondary and the primary masses.

The minimum separation of the binary and, assuming circular 
Keplerian orbits, the minimum value of $\tau_{\rm orb}$ for which both SPT and FUVT are applicable is obtained by equating eq.~\ref{eq:bentz2} and eq.~\ref{eq:RL}:
\begin{eqnarray}\label{eq:mintau}
    \tau_{\rm orb, min} & \approx & 200 \, {\rm yr} \, f_{\rm Edd, 2}^{0.78}\, \left(\frac{M_2}{10^6\,{\rm M}_\odot}\right)^{0.28} \nonumber \\ 
     & & \sqrt{\frac{[0.6\, q^{2/3}\, + \, \ln (1+q^{1/3})]^3}{q \, (1+q)}},
\end{eqnarray}
where $f_{\rm Edd, 2}$ is the Eddington ratio of the secondary.
It is interesting to note that $\tau_{\rm orb, min}$ has a stronger dependence on $f_{\rm Edd, 2}$ than on $M_2$. Therefore, at fixed luminosity, shorter minimum periods are expected for higher MBH  masses. Similarly to the spectroscopic MBHB candidates proposed to date \citep[e.g.][]{Tsalmantza11,Eracleous12}, secondary MBHs of $10^8$~M$_\odot$ could be in binaries with minimum periods of as small as $\sim 27$ yr \citep[see also][]{decarli13, runnoe17} if accreting at $f_{\rm Edd, 2} =0.01$, that is, with the same bolometric luminosity as an MBH of $10^6$~M$_\odot$ accreting at its Eddington limit. 

However, we note that for disc-like BLRs, the region within which circular orbits around a single MBH are stable is sizeably smaller (by a factor of $\approx 4-5$) than the Roche lobe \citep[e.g.][]{eggleton83,runnoe15size}. On top of this consideration, gas at distances of $>R_{\rm B-H\beta}$ contribute to the central part of the broad H$\beta$ line, and, if the binary separation is too small, such a contribution would be lost and the line would acquire a `boxy' profile\footnote{A boxy profile is not an unequivocal signature of a binary. Indeed, some of the observed AGN discussed in section 3 have boxy profiles; see e.g. the mean spectra in \cite{bentz09a}.}. In the following, we  consider binaries with periods corresponding to BLR sizes similar to the secondary Roche lobe, as well as binaries with significantly larger periods, for which a stable disc-like BLR can survive around each of the two MBHs (see section 4).

Finally, the typical timescale for the H$\beta$ reverberation to the change of the continuum can be estimated as $\tau_{\rm B-H\beta}=R_{\rm B-H\beta}/c$ (where $c$ is the speed of light). This timescale was evaluated using RM and ranges from days to months depending on the luminosity of the accreting MBH. However, RM studies typically require longer durations ---even for $\sim 10^6$ M$_\odot$ MBHs--- due to the need for sufficiently long baselines and sufficiently varying continua to allow for the test. As a consequence, the typical timescales needed to probe the variability both in the continuum and in the BELs are of the order of weeks to months \citep{bentz09a}.

\section{FUVT description and results on a control sample of single MBHs}\label{sec:single}

We apply FUVT to nine AGN for which $\tau_{\rm B-H\beta}$ (and therefore masses) were obtained by \cite{bentz09a}: Mrk 142, SBS 1116+583A, Arp 151, Mrk 1310, Mrk202, NGC 4253, NGC 4748, NGC 5548, and NGC 6814. We assume, as a working hypothesis, that the whole sample is composed of only bona fide single MBHs, and use it as a comparison for the mock binary sample described in the following section. Nevertheless, we stress that at least one of the AGN in sample I (NGC 5548) has been proposed as a binary candidate \citep{Li2016} based on the variability of the periodic modulation of its broad H$\beta$ line.

For all AGN, \cite{bentz09a} estimated delays between the continuum and the broad H$\beta$ of between 2 and 7 days and masses in the range of $1-7 \times 10^6$ \msun, with the exception of NGC 6814, which has a mass $\approx 1.85 \times 10^7$ \msun, and NGC 5548, which has a mass of $8.2 \times 10^7$ \msun \citep[see][for additional details on the properties of the single AGNs]{bentz09a}.

For our analysis, we used scaled spectra released by these latter authors after the application of a renormalisation used to set the flux of all spectra to a consistent scale. For every AGN, we removed the mean spectrum (averaged over all the observations) and worked on the variable part of the spectrum only, which is dominated by the AGN continua and the BELs. We then removed the AGN continuum by fitting a straight line to the spectrum near the H$\beta$ BEL using the same wavelength intervals as those used in \cite{bentz09a}.\footnote{Although our procedure is slightly different from that used in \cite{bentz09a}, we recover the same fluxes (within uncertainties) from the whole broad H$\beta$  for every pointing of every AGN.}

For each AGN, we then computed the root mean square (RMS) spectrum. We used it to identify seven $\bar{\lambda}$ dividing the RMS broad 
H$\beta$ into eight  parts of equal flux\footnote{The number of $\bar{\lambda}$ used to spit the BEL is indeed somewhat arbitrary. We used eight as it is the maximum number of bins used in \cite{bentz09a} for 2D RM of the nine single MBHs, which was chosen in order to keep a sufficiently high signal-to-noise ratio in each frequency bin.}. 
For each $\bar{\lambda}$ in each spectrum, we divided the H$\beta$ BEL into two components, a red one and a blue one (r-H$\beta$ and b-H$\beta$, for wavelengths longer and shorter than $\bar{\lambda}$, respectively). We then computed the light curves of the r-H$\beta$ and b-H$\beta$ components, and performed the cross-correlation between the two,
allowing for a small time-shift $\tau_{\rm shift}$ in the [$-10$ day, $10$ day] interval to take into account different reverberation times for different parts of the broad lines \citep[due to e.g. the inflowing or outflowing BLR dynamics; e.g.][]{bentz09a}. More specifically, we follow the standard RM practice of first using one light curve and interpolating the other, before swapping the two and repeating the exercise. The average cross-correlation is then used, and the uncertainties on the the cross-correlation and $\tau_{\rm shift}$ are estimated using the  public \textsc{Python}  version \citep{python18}\footnote{We modified the public version to also output the uncertainties on the cross-correlation.} of the Monte Carlo code discussed in \cite{peterson98, peterson04}.

\begin{figure}
    \centering
    \includegraphics[width=0.48\textwidth]{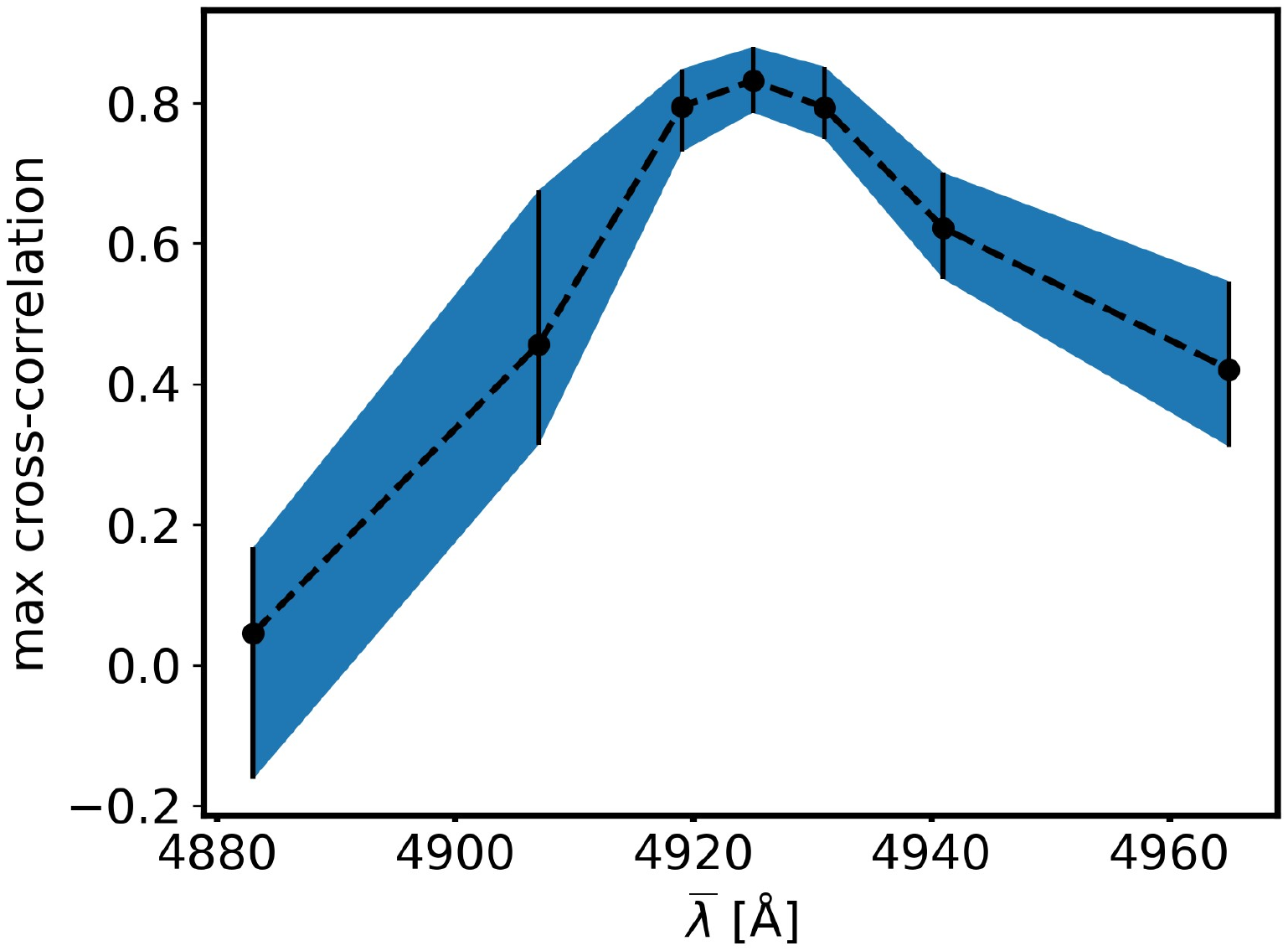}
    \caption{NGC 4748: Maximum value of the cross correlation between the blue and red parts of the broad H$\beta$ line as a function of the dividing wavelength $\bar{\lambda}$. The peak of the cross-correlation is obtained when the line is divided into two parts of equal flux, which is when the signal-to-noise ratio is the highest for both the red and blue sides.}
    \label{fig:CCF1}
\end{figure}

For each AGN and each $\bar{\lambda,}$ we measure $max-CCF(\bar{\lambda})$, that is, the maximum value of the r-H$\beta$--b-H$\beta$ cross-correlation, and its uncertainties. The $max-CCF(\bar{\lambda})$  obtained following this procedure are shown in figure~\ref{fig:CCF1} for NGC 4748. The peak of $max-CCF(\bar{\lambda})$ ($\overline{CCF}$ hereafter)\footnote{We stress again that $\overline{CCF}$ refers to the maximum CCF over all the possible $\bar{\lambda,}$ while $max-CCF(\bar{\lambda})$ refers to the maximum CCF for a specific $\bar{\lambda}$.} corresponds to the $\bar{\lambda}$ threshold dividing the broad H$\beta$ line into two parts of  equal flux, while the cross-correlation decreases significantly (and its uncertainties increase) when moving $\bar{\lambda}$ towards the line wings, which are more affected by the noise in the spectra.
More generally, in the sample, the values of $\overline{CCF}$ are comparable to or slightly higher than the peak of the cross-correlations between the whole broad H$\beta$ and the photometric B and V  light curves shown in \cite{bentz09a}. The trends discussed for NGC 4748 are common for all nine of the AGN we examined:  $\overline{CCF}$ is always found in the bulk of the line (hereby defined as the region in between the third and the fifth values of $\bar{\lambda}$, i.e. when the least luminous part of the BEL has at least three-eighths of the total BEL flux), and it is always $\gsim 0.75$.
\footnote{The profiles of $max-CCF(\bar{\lambda})$ for all the other single MBHs as well as for all the mock binaries discussed in the following section are available at https://astro.fisica.unimib.it/spectroscopical-search-of-massive-black-hole-binaries/}.

\section{Application of FUVT to mock MBHBs}\label{sec:mock}

\begin{figure}
    \centering
    \includegraphics[width=0.48\textwidth]{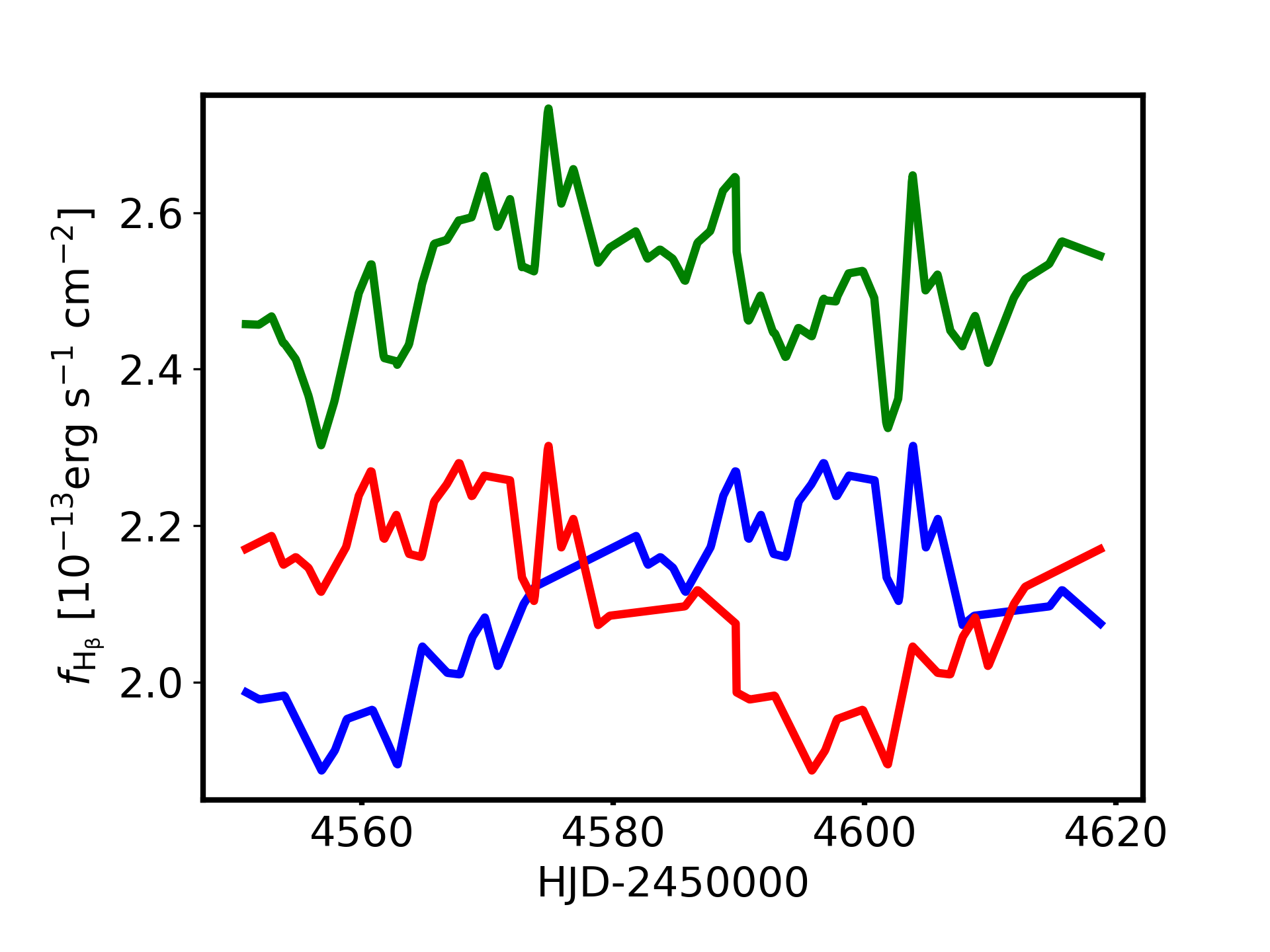}
    \caption{NGC 4748: Original light curve of the entire H$\beta$ line (blue line), the same light curve shifted by half of the observational period (with periodic boundary conditions, red line), and the sum of the two (rigidly shifted towards lower fluxes for visualisation purposes, green line), mimicking a MBHB whose components evolve independently on short timescales.}
    \label{fig:lightcurves}
\end{figure}

We apply FUVT to mock equal-mass MBHBs generated by duplicating the temporal series of spectra of each of the nine single MBHs discussed above.
One of the copies is shifted in time  by half of the observational period (assuming periodic boundary conditions, as done in standard reverberation mapping studies). As the magnitude of the time-shift $\tau_{\rm shift}$ in the test is constrained to be smaller than 10 days, such a shift makes the evolution of the two series of spectra independent. The original and shifted light curves (blue and red lines) of the whole H$\beta$ for NGC 4748, together with the sum of the two (green line), mimicking the light curve of a MBHB, are shown in figure~\ref{fig:lightcurves}.\footnote{We acknowledge that we could also enforce not-causally related light curves by using two different AGN. Unfortunately, due to the limited number of single MBHs, only two pairs have a small enough redshift difference to be interpreted as a relative velocity between two loosely bound MBHs. We defer an investigation of `heterogeneous' mock binaries to a future study.} 

The two series of spectra are then shifted in frequency by the same amount towards higher and lower wavelengths, respectively. The wavelength shifts are chosen in order to mimic circular MBHBs with periods of $\tau_{\rm orb} = 50, 100, 300,$ and 1000 yr, observed close to edge-on at the orbital phase that would maximise the Doppler effect. We chose such a configuration to check whether or not  the test succeeds in identifying binaries in the most favourable scenario. Different configurations would result in smaller shifts between the two independent components of the BEL, and could in principle result in a missed detection. We note that, while all four periods fulfill the criterion set in equation~4, $\tau_{\rm orb} = 50, 100$ would correspond to separations at which a sizable part of a disc-like BLR would be unstable. Furthermore, for those periods, the outer parts of the BLR would be removed, resulting in a more `boxy' profile of each BEL component \citep[see e.g.][]{nguyen19}, while only a few of the AGN discussed in section~3 have a boxy broad H$\beta$. For these reasons, hereafter we refer to binaries with $\tau_{\rm orb} = 300$ and 1000 yr as `solid' binaries, while keeping the $\tau_{\rm orb} = 50$ and 100 yr binaries to highlight how the performance of our test scales with the orbital period and to account for possible alternative geometries and dynamics of the BLR.

\begin{figure}
    \centering
    \includegraphics[width=0.48\textwidth]{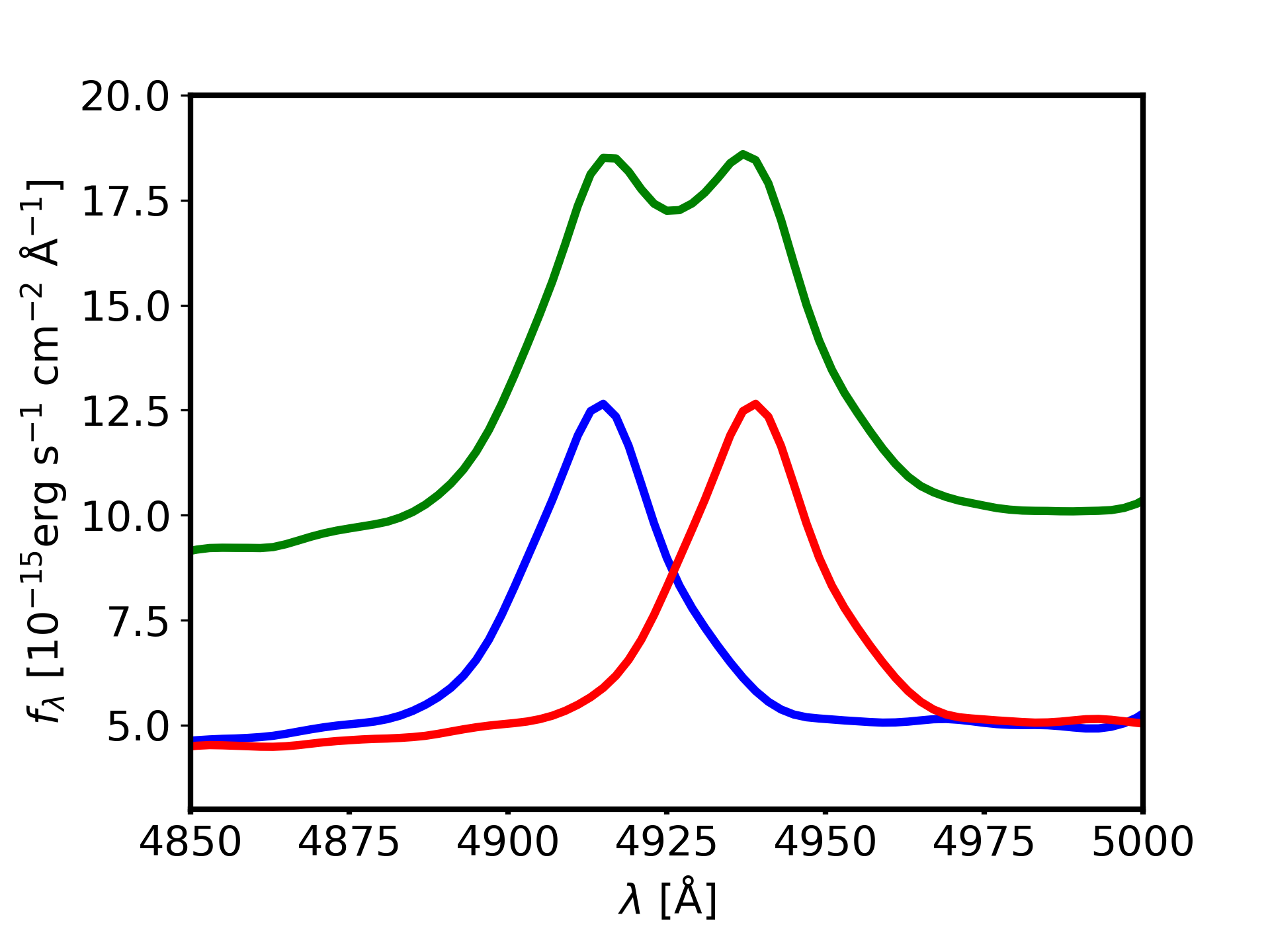}
    \includegraphics[width=0.48\textwidth]{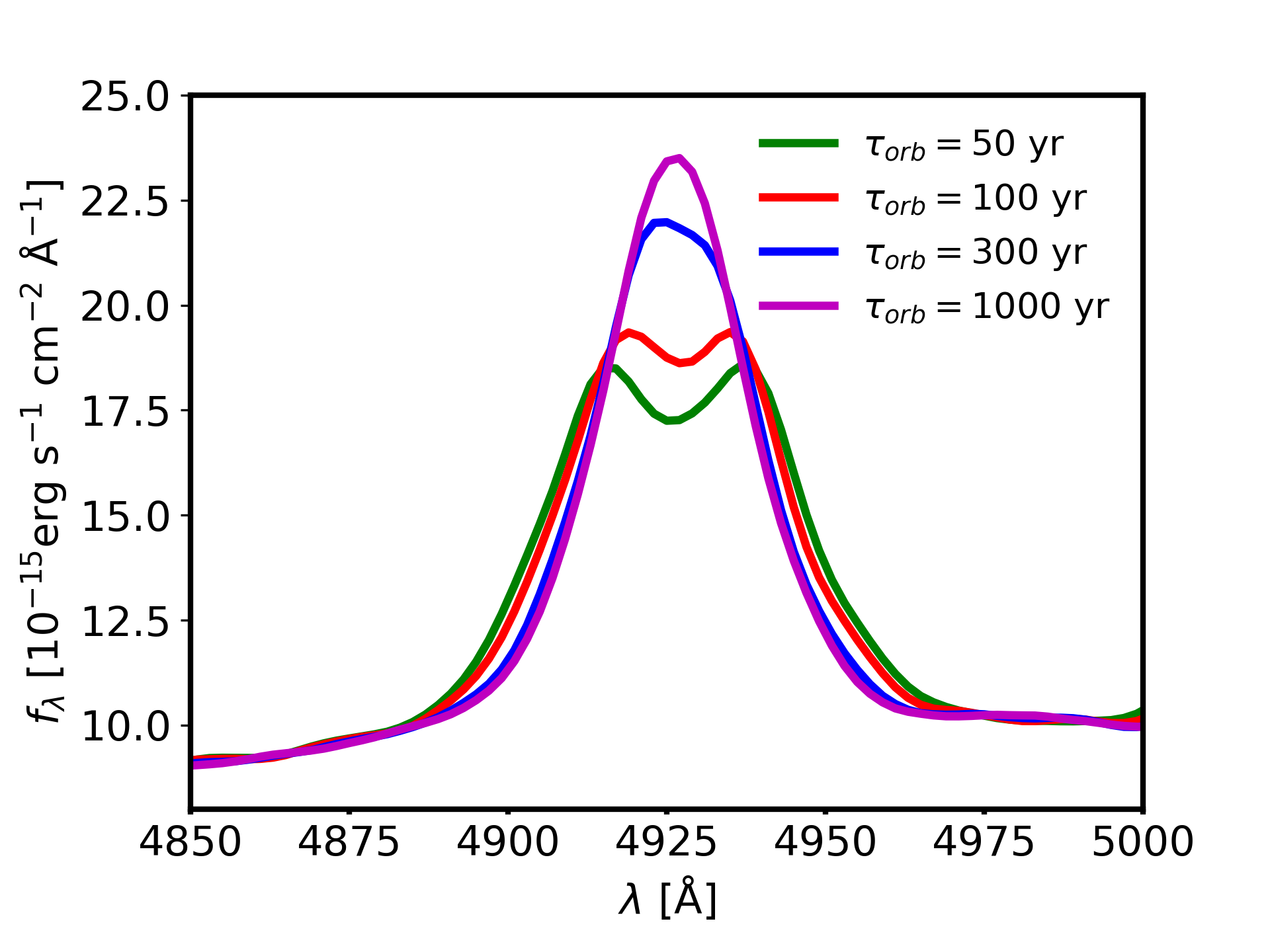}
    \caption{NGC 4748. Upper panel: Broad H$\beta$ component of  a mock spectrum shifted redwards (red line) and bluewards (blue line) to mimic the orbital velocity of an equal-mass MBHB with orbital period $\tau_{\rm orb} = 50$ yr. The green line shows the composite line associated to the mock MBHB.
    Lower panel: Composite broad H$\beta$ lines for mock MBHBs with $\tau_{\rm orb} = 50, 100, 300,$ and 1000 yr (green, red, blue, and magenta lines, respectively).
    The Doppler shifts are evaluated assuming an edge-on circular binary with the MBH velocities  perfectly aligned with the line of sight.}
    \label{fig:shift}
\end{figure}

We then add the two series up (adding the original errors in quadrature for each wavelength bin), obtaining four new mock MBHBs for each original MBH, for a grand total of 36 mock MBHBs. The upper panel of figure~\ref{fig:shift} shows an example of a mock MBHB spectrum based on the observations of NGC 4748 ---without the application of any time shift, for clarity purposes--- and for a binary with $\tau_{\rm orb} = 50$ yr. In the lower panel the resulting line profile starting from the same spectrum are shown for four different  mock binary periods. 

\begin{figure}
    \centering
    \includegraphics[width=0.48\textwidth]{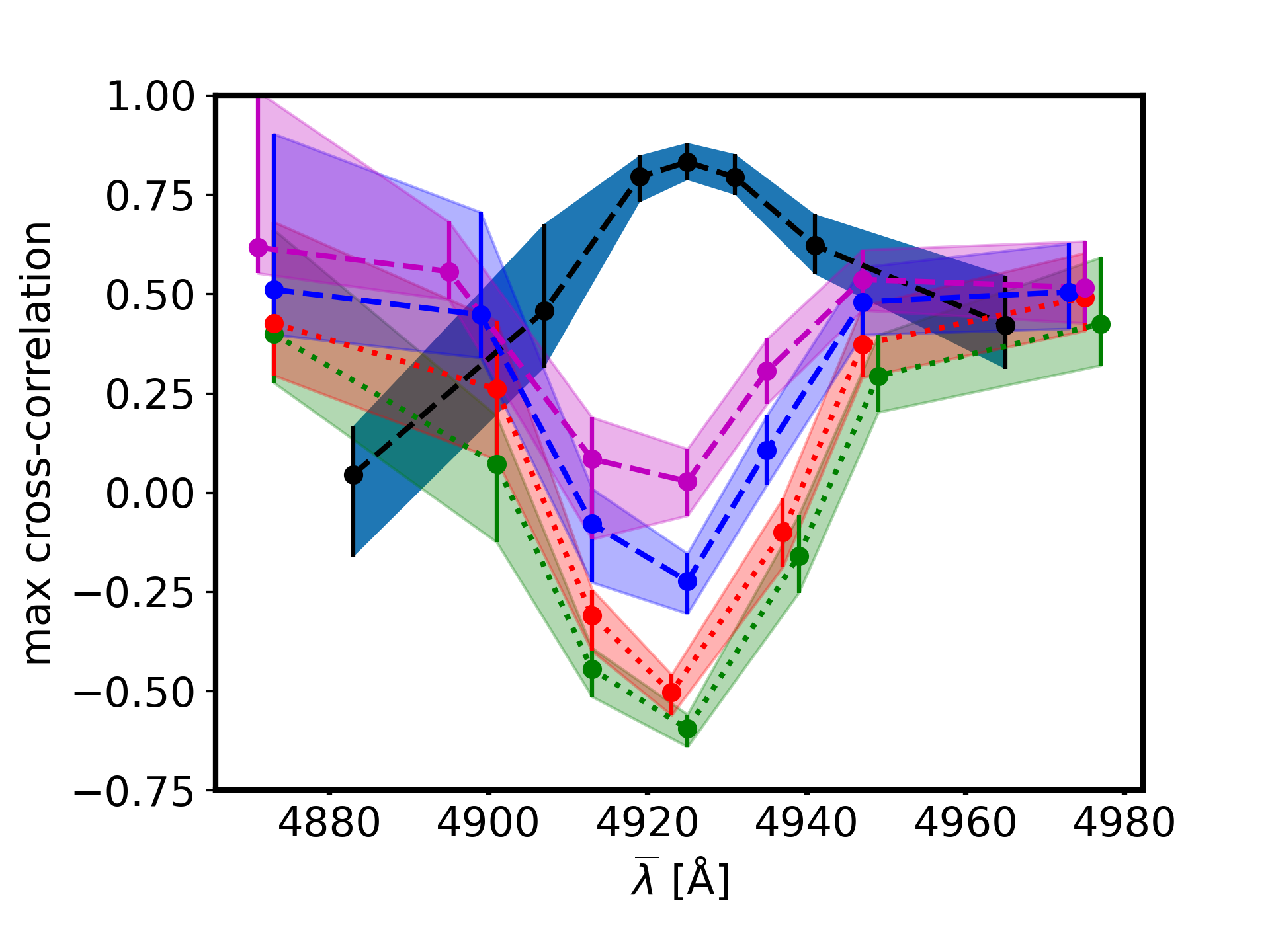}
    \caption{Same as figure~\ref{fig:CCF1}, but for the four mock MBHBs obtained from NGC 4748. The green, red, cyan, and magenta points with uncertainties are for binaries with $\tau_{\rm orb} = 50, 100, 300,$ and 1000 yr, respectively. The original values and uncertainties of NGC 4748 are shown in blue for comparison. `Solid' binaries are shown with dashed lines, while dotted lines show shorter-period binaries.}
    \label{fig:binaryccf}
\end{figure}

We then apply FUVT to  these new sets of mock spectra.
The resulting profile of $max-CCF(\bar{\lambda})$  as a function of $\bar{\lambda}$ for the four mock MBHBs constructed starting from NGC 4748 are shown in figure~\ref{fig:binaryccf}.

The four mock binaries  show clearly different $max-CCF(\bar{\lambda})$ profiles with respect to the single MBH case; they tend to have significantly smaller values of $max-CCF(\bar{\lambda})$  than the single MBH cases close to the bulk of the broad line: the shorter the period, the larger the frequency separation between the binary component contributions, and the lower the $max-CCF(\bar{\lambda})$. The highest values of the mock MBHB cross-correlation are instead found close to the line wings, where a correlation might be found due to the low signal-to-noise ratio of one of the two sides of the line. The anti-correlation observable for central values of  $\bar{\lambda}$ is due to the shape of the broad H$\beta$ light curve, which features a single peak appearing after a dimmer state (see figure~\ref{fig:lightcurves} and the following discussion in this section). Such an anti-correlation should not be considered a solid feature associated with binarity; indeed, it is not observed in some of the other systems. We note that even the mock binary with a 1000 yr period shows a clear minimum in the  $max-CCF(\bar{\lambda})$ profile, with the two parts of the line being almost completely uncorrelated, regardless of the relatively small velocity shift between the two components associated with the two MBHs.
In this case, the relative velocity between
the two MBHs is $\approx 500$ km$/$s (equivalently, the shift of one of the
peaks with respect to the centroid is $\approx$250 km$/$s), and the FWHM of the H$\beta$ line is
$\approx$2000 km$/$s ($\sigma \approx$ 1000 km$/$s) in the mean spectrum for NGC4748. If we consider the varying component only, looking at the RMS
spectrum for the same system, we get FWHM$\approx$1200 km$/$s ($\sigma \approx$650 km$/$s)\footnote{All the values of $\sigma$ and FWHM obtained from our analysis are consistent with those presented in \citep{bentz09a}.}. While the velocity shift is smaller than the FWHM, our test still manages to identify the mock as a binary.

\begin{figure*}
    \centering
    \includegraphics[width=0.48\textwidth]{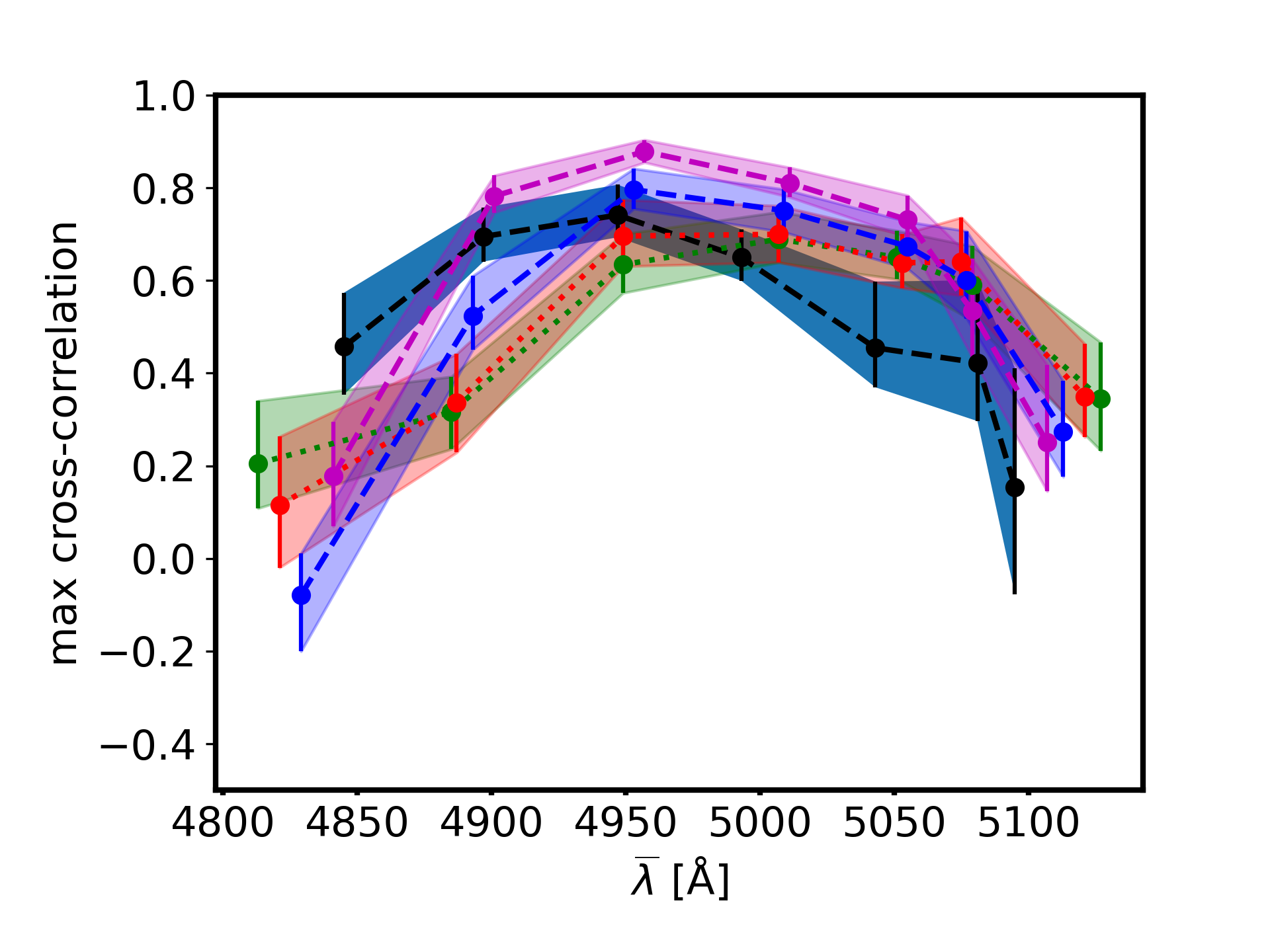}
    \includegraphics[width=0.48\textwidth]{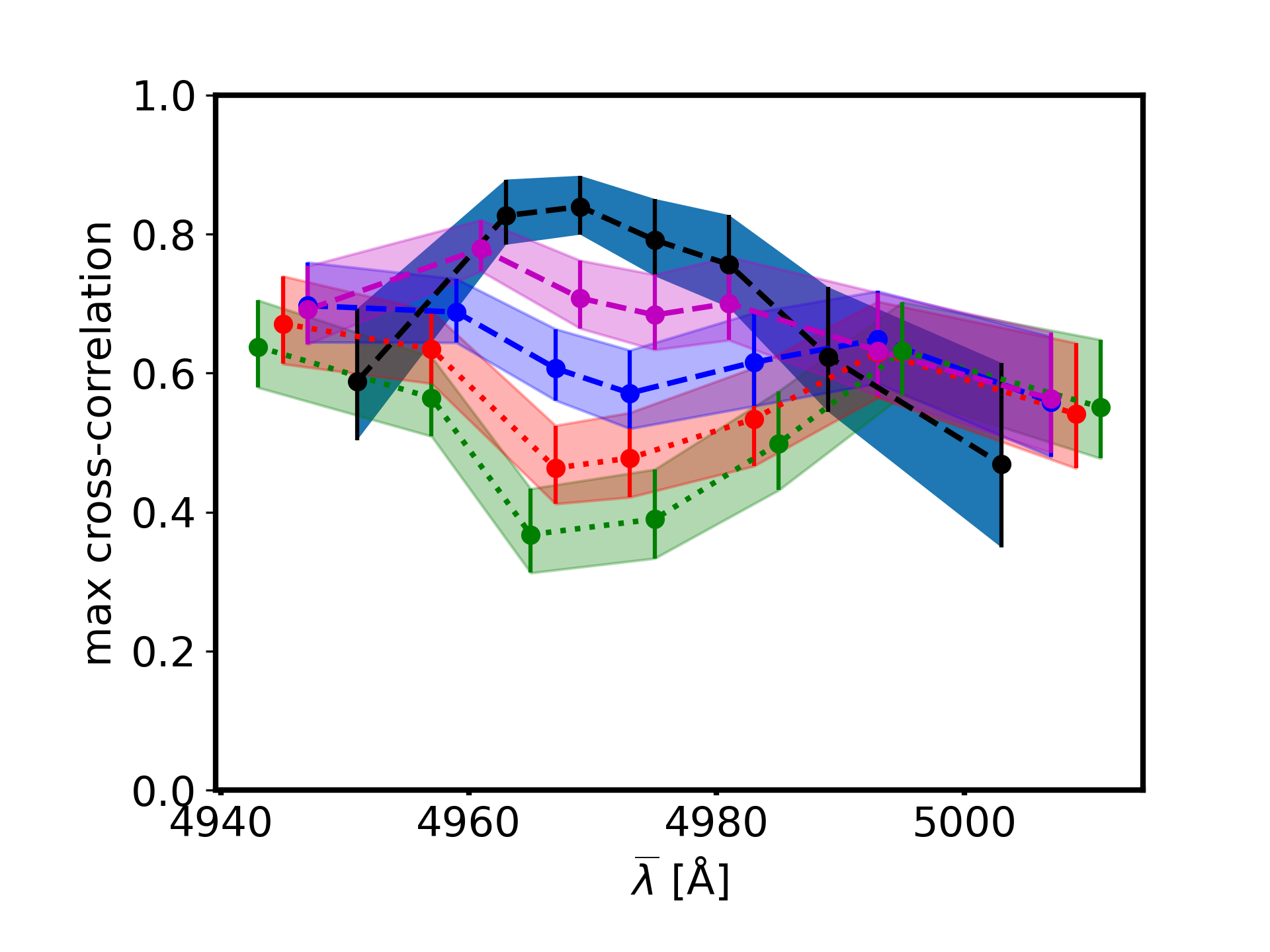}
    \caption{Same as figure~\ref{fig:binaryccf} but for NGC 5548 and the associated mock binaries (left panel) and for Mrk 202 (right panel). The case on the left represents a class of objects in which the max cross-correlation in the bulk of the broad line can be higher or lower than that of the single MBH depending on the assumed period. The right panel shows the only case in which the single MBH  has a value of $\overline{CCF}$ of higher than any of its sibling mock binaries, but the cross-correlation of the mocks can be quite high, up to $\sim 0.75$ for the shortest period. Please note that the y axis changes from panel to panel in order to emphasise the differences between the different cases.} 
    \label{fig:binaryccf1}
\end{figure*}

More generally, four other sets of mock binaries based on other observed AGN show trends that are qualitatively similar to the one shown in figure~\ref{fig:binaryccf}; three sets have values of $\overline{CCF}$ that can be higher or lower than that of the single MBH case, depending on the choice of orbital period; and one set has lower $\overline{CCF}$ compared to the corresponding single MBH, which nevertheless remains relatively high, up to $\approx 0.75$. An example mock binary for each of the last two classes is shown in figure~\ref{fig:binaryccf1}\footnote{The profiles of $max-CCF(\bar{\lambda})$ for all the other single MBHs and mock binaries are available at https://astro.fisica.unimib.it/spectroscopical-search-of-massive-black-hole-binaries/.}.
The occurrence of different behaviours is not surprising, because the actual profile of $max-CCF(\bar{\lambda})$ depends mostly on the shape of the original H$\beta$ light curve used to generate the mock binary. If the original H$\beta$ light curve displays a single peak and single drop with similar duration, the corresponding binaries will tend towards anti-correlation when two-close-to-equal-flux parts of the composite H$\beta$ are considered. If the original H$\beta$ light curve instead shows a more rapid variability, the shift of half of the duration of the observations may result in close to zero correlation. If instead the H$\beta$ light curve, for example, shows two peaks and two troughs, both equally spaced and of similar duration, the time shift will not significantly reduce the maximum cross-correlation. We stress that such a variety of behaviours is due to the limited timescale of the spectroscopic monitoring campaigns. In principle, in the binary scenario, a campaign of arbitrary timescale would have a collection of BEL profiles (one per pointing) that are contributed from the two completely uncorrelated components (1 and 2) associated to the two MBHs. In this case, when cross-correlating the blue and red parts of the total BEL, the cross terms in the cross-correlation would be equal to zero \footnote{I.e. the red part of component 1 ($r_1$) would have a zero cross-correlation with the blue part of component 2 ($b_2$).}. 
The cross correlation between the two sides of the BEL would then read:
\begin{eqnarray}\label{eq:anal_est_1}
CCF(r,b,\tau)&=CCF(r_1 + r_2, b_1 + b_2, \tau)\nonumber \\ 
&= \frac{\sigma(r_1) \, \sigma(b_1)}{\sigma(r_1 + r_2) \, \sigma(b_1 + b_2)} CCF(r_1,b_1,\tau)\nonumber \\ 
&+\frac{\sigma(r_2) \, \sigma(b_2)}{\sigma(r_1 + r_2) \, \sigma(b_1 + b_2)} CCF(r_2,b_2,\tau),
\end{eqnarray}
where $\sigma(x)$ is the standard deviation of the $x$ quantity. In the simplifying case, in which the time delay maximising the cross-correlation is the same for the two components, as is the case for our mock binaries, we obtain
\begin{eqnarray}\label{eq:anal_est_2}
CCF(r, b, \tau_{\rm max}) = \frac{\sigma(r_1) \, \sigma(b_1)}{\sigma(r_1 + r_2) \, \sigma(b_1 + b_2)} CCF(r_1,b_1,\tau_{\rm max})\nonumber \\ 
+\frac{\sigma(r_2) \, \sigma(b_2)}{\sigma(r_1 + r_2) \, \sigma(b_1 + b_2)} CCF(r_2,b_2,\tau_{\rm max}).
\end{eqnarray}
From eqs.~\ref{eq:anal_est_1} and \ref{eq:anal_est_2}, we can derive some interesting trends and limits: when the shifts of the two components (due to the velocity of the two MBHs along the line of side) tend to zero, and the line is split approximately in half, the denominator of the $\sigma$ ratios tends to $2\, \sigma(r_1)\,\sigma(b_1)\approx 2\, \sigma(r_2)\,\sigma(b_2)$, and the cross-correlation in eqs.~\ref{eq:anal_est_1} and~\ref{eq:anal_est_2} tends to the arithmetic average of the cross-correlation between the blue and red parts of the two single components. If these cross-correlations are strong (as expected for a single MBH--BLR system; see section~\ref{sec:single}), the total cross-correlation will be equally strong, as expected when there is zero velocity shift between the two components. On the other hand, the maximum cross-correlation in eq.~\ref{eq:anal_est_2}  will become lower and lower (while remaining positive) as the two components 1 and 2 become more shifted from the reference frame of the galaxy and contribute more asymmetrically to the red and blue parts of the whole line. Unfortunately, in many cases (such as those considered in this paper), the length of the RM campaigns will not be sufficient to ensure that the two components are completely uncorrelated, leading to the variety of outcomes mentioned above. The dependence of the profile of $max-CCF(\bar{\lambda})$ for binaries on the specific realisation of the variability observed during the RM campaign has important implications for the identification of the studied systems as MBHBs. In following section, we further discuss this point.

\begin{figure}
    \centering
    \includegraphics[width=0.48\textwidth]{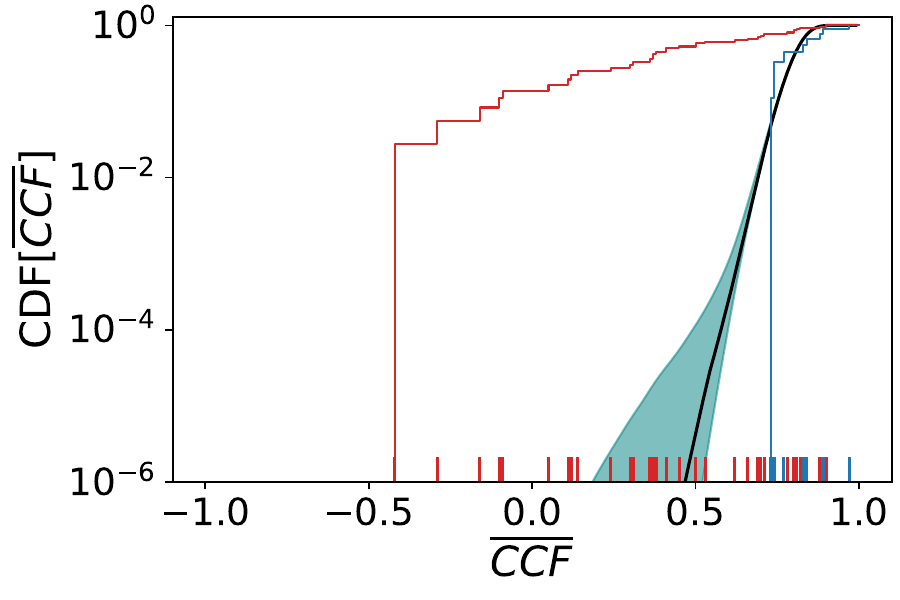}
    \caption{Normalised cumulative distribution of the observed single MBHs (blue histogram) and of our mock MBHBs (red histrogram). The solid black line shows the median cumulative probability that a system will have a $\overline{CCF}$ of lower than a given value (obtained from the single MBH data under the assumptions specified in the main text). The teal band indicates the 90\% credible region for the cumulative distribution of $\overline{CCF}$ for single MBHs. The vertical red (blue) ticks highlight the $\overline{CCF}$ values for all of the 36 mock binaries (9 single MBHs). The numerical values of all of the $\overline{CCF}$ and cumulative probabilities  are listed in table~\ref{tab:prob}. 
    }
    \label{fig:prob}
\end{figure}

In principle, with a sufficiently large population of single MBHs studied with RM in order to constrain the distribution of $\overline{CCF}$ 
in the line bulk (i.e. measured for one of the three central $\bar{\lambda}$), we could translate the values of $\overline{CCF}$ of our mock binaries into a probability of not belonging to the `standard' population. To date, this distribution is largely unconstrained. As an example, in figure~\ref{fig:prob} we show the distributions of $\overline{CCF}$ for the observed single MBHs discussed in section~\ref{sec:single} (blue histogram) and the mock binaries (red histograms). The two distributions are clearly different. Starting from the observed $\overline{CCF}$ for the nine AGNs, we construct a model of the $\overline{CCF}$ distribution for `typical' single MBH, which we then use to infer the probability that mocks belong to the single MBH population. Our model is based on the assumption that ($i$) the population of single MBHs has a well-defined value of $\overline{CCF}$, and that ($ii$) the red and blue parts of the H$\beta$ flux data follow a bivariate Gaussian distribution, with the aforementioned, unknown correlation coefficient $\overline{CCF}$. The latter allows us to exploit the analytical form of the sampling distribution of the $\overline{CCF}$ derived by \citet{fisher28} given a set of observed $\overline{CCF}_i$. We infer a posterior distribution for $\overline{CCF}$ using a nested sampling algorithm, which we use as the model for the expected distribution of measured  $\overline{CCF}$ for single MBHs. The cumulative probability of observing a given value of $\overline{CCF}$ (again restricting our search to the bulk of the BELs) is shown as a black curve in figure~\ref{fig:prob}.
The vertical blue and red ticks highlight the values of $\overline{CCF}$ measured for all the observed single and mock binaries in the bulk of the BEL. Mocks with lower values of $\overline{CCF}$ have lower probabilities, with $24$ (out of 36) mocks having a probability of $< 10^{-3}$. The values of $\overline{CCF}$ and of the associated probabilities are reported in Table~\ref{tab:prob}.\footnote{We stress that the somewhat low probability values for two of the single MBHs (Mrk142 and NGC5548) should be considered with caution, because the functional form of the probability distribution was decided a priori for this exercise.}

\begin{table}
        \centering
    \caption{Values of $\overline{CCF}$ and the associated probability of belonging to the distribution of single MBHs for single and (mock) binary MBH.}
        \label{tab:prob}
        \begin{tabular}{lccr} 
                \hline
                name & $\overline{CCF}$ & P & `solidity'\\
                \hline
                Mrk142 & 0.73 & 0.04 &\\
        SBS1116+583A & 0.77 & 0.2 &\\
        Arp151 & 0.97 & $>0.99$ &\\
        Mrk1310 & 0.89 & $>0.99$&\\
        Mrk202 & 0.84 & 0.71 &\\
        NGC4253 & 0.74 & 0.06&\\
        NGC4748 & 0.83 & 0.64 &\\
        NGC5548 & 0.74 & 0.06&\\
        NGC6814 & 0.88 & 0.97&\\
                \hline
                Mrk142\_50 & 0.66 & $<10^{-3}$&\\
        Mrk142\_100 & 0.71 & 0.01&\\
        Mrk142\_300 & 0.78 & 0.26&solid\\
        Mrk142\_1000 & 0.81 & 0.49&solid\\
        SBS1116+583A\_50 & 0.38& $<10^{-3}$&\\
        SBS1116+583A\_100 & 0.37& $<10^{-3}$&\\
        SBS1116+583A\_300 & 0.41& $<10^{-3}$&solid\\
        SBS1116+583A\_1000 & 0.45& $<10^{-3}$&solid\\
        Arp151\_50 & -0.42 & $<10^{-3}$&\\
        Arp151\_100 & -0.29 & $<10^{-3}$&\\
        Arp151\_300 & -0.09 & $<10^{-3}$&solid\\
        Arp151\_1000 & 0.14 & $<10^{-3}$&solid\\
        Mrk1310\_50 & 0.05 & $<10^{-3}$&\\
        Mrk1310\_100 & 0.12 & $<10^{-3}$&\\
        Mrk1310\_300 & 0.24 & $<10^{-3}$&solid\\
        Mrk1310\_1000 & 0.41 & $<10^{-3}$&solid\\
        Mrk202\_50 & 0.5 & $<10^{-3}$&\\
        Mrk202\_100 & 0.53 & $<10^{-3}$&\\
        Mrk202\_300 & 0.62 & $<10^{-3}$&solid\\
        Mrk202\_1000 & 0.71 & 0.01&solid\\
        NGC4253\_50 & 0.3& $<10^{-3}$&\\
        NGC4253\_100 & 0.37& $<10^{-3}$&\\
        NGC4253\_300 & 0.36& $<10^{-3}$&solid\\
        NGC4253\_1000 & 0.5 & $<10^{-3}$&solid\\
        NGC4748\_50 & -0.16& $<10^{-3}$&\\
        NGC4748\_100 & -0.1& $<10^{-3}$&\\
        NGC4748\_300 & 0.11& $<10^{-3}$&solid\\
        NGC4748\_1000 & 0.31& $<10^{-3}$&solid\\
        NGC5548\_50 & 0.69 & 0.004&\\
        NGC5548\_100 & 0.7 & 0.008&\\
        NGC5548\_300 & 0.8 & 0.4& solid\\
        NGC5548\_1000 & 0.88 & 0.97& solid\\
        NGC6814\_50 & 0.8& 0.41&\\
        NGC6814\_100 & 0.82& 0.56&\\
        NGC6814\_300 & 0.9& $>0.99$& solid \\
        NGC6814\_1000 & 0.9& $>0.99$& solid\\
                \hline
        \end{tabular}
        \tablefoot{The values of $\overline{CCF}$ are restricted to the bulk of the line, while the associated probability of belonging to the distribution of single MBHs have been calculated using the median cumulative distribution for $\overline{CCF}$ shown with the solid line in Fig.~\ref{fig:prob}. The names of the mocks are constructed by attaching the period of the binary (in yr) to the name of the real AGN progenitor. The last column identifies the `solid' binaries, for which a stable disc-like BLR can exist around both MBHs. Considering that our inferred cumulative distribution for the $\overline{CCF}$ is well determined up to probabilities of as low as $\sim 10^3$, we consider smaller values as upper limits.}
\end{table}

\section{Discussion}\label{sec:discussion}

We present a novel method (FUVT) to search for MBHBs and to test the binary hypothesis in already identified MBHB candidates. The method is tailored to seek large separation binaries, when the two MBHs can retain their own BLR, with orbital periods that can be $\gsim 50$ yr. 
FUVT assumes that the two MBHs are at sufficiently large distances to ensure that the short-term variability of the two BLRs is uncorrelated. A simple correlation test between the long- and short-wavelength parts of BELs can therefore identify binaries if the maximum cross-correlation (once the BELs have been split into two comparable flux components) is sufficiently small compared to the high level of correlation expected and observed for single MBHs.

As opposed to the already proposed SPT, which  requires follow-up campaigns with a minimum duration of $\sim \tau_{\rm orb}$  in order to track the expected evolution of the orbital velocity of one active component of the binary, FUVT works on significantly shorter observational campaigns. The expected timescales of the test are from weeks to a few years in the worst-case scenario, provided that a sufficiently frequent time coverage (typical of RM studies) of the candidate spectra is achieved.

In addition to the obvious advantage of its short duration, FUVT has a number of specific advantages and disadvantages with respect to SPT, making the two procedures complementary:
\begin{itemize}
    \item FUVT performs best when the two MBHs in the binary have  similar BEL luminosities, on average. According to theoretical studies \citep[e.g.][and references therein]{roedig12,duffell20}, secondary MBHs are expected to experience higher accretion rates, making FUVT particularly relevant either for  systems of close
to equal mass (and equal Eddington ratio)  or for systems in which the secondary luminosity is close to its Eddington limit. If the flux coming from one of the two BLRs is too small compared to the other, FUVT could fail in distinguishing the subdominant contribution from the observational noise.
    \item If however the two MBHs contribute similarly to the BELs, the previously proposed tests (and SPT in particular) could miss real binaries, because at large separations, double-peaked profiles are not always  detectable (see e.g. fig~\ref{fig:shift}) or could be misinterpreted as the indication  of a disc-like BLR, as in the standard interpretation for the broad double-peaked emitters \citep[e.g.][]{eracleous94}. Even if a binary is selected as a candidate at a time when one MBH was significantly brighter than the other, SPT could fail in observing a long-term frequency shift consistent with the orbital evolution of a binary if the other component undergoes a sizable rebrightening, changing the shape (and centroid) of the line. Such an evolution could result in the erroneous dismissal of real binaries if only SPT is performed, while the binary nature of the system would be easily identifiable by FUVT.
    \item As discussed in section~\ref{sec:mock}, the stochastic nature of the short-term variability of each MBH accretion disc can occasionally result in strong cross-correlations even if the observed AGN hosts a real MBHB. If FUVT is applied to large spectroscopic catalogues to identify new MBHB candidates, a fraction of the MBHBs in the data are expected not to be missed. 
    It is difficult to estimate a solid missed detection fraction because of the small number of single MBHs that we use to characterize the `control sample'. However, taking the probabilities we estimated (with a priori assumptions on the shape of p(CC), which will be verified when enough data become available), and assuming a `missed' alarm  threshold of $p \geq 0.01$ (corresponding to $CC \geq 0.70$), we get 10 systems out of 36 ($\approx 28\%$ of missed binaries), including binaries with a period shorter than 300 yr, or 9 out of 18 when considering only larger `solid' binaries (see discussion in section 2). Such shortcomings can be circumvented when FUVT is  applied to data taken from a significantly longer campaign with equally frequent observations, motivated for example by the objective to independently identify a source as a promising MBHB candidate. In this case, if the first observational campaign finds a high correlation in the red and blue part of the BELs, the test should be considered inconclusive, and a new campaign should be performed. Only after a few campaigns (for a total of a few years, in the worst-case scenario) can the binary hypothesis be ruled out (unless a low correlation period is found, in which case the binary nature is indeed confirmed). However, we note that the missed detection fraction estimated here cannot be used to infer the statistics of the global MBHB population, because in this first method paper we are considering only the idealised edge-on and Doppler-maximising configuration for equal-mass binaries. A broader analysis of the binary parameter space, including a statistical estimate of the minimum number of observational campaigns needed to disprove the MBHB scenario with at a given confidence level, is postponed to a future study.
    \item Finally, we stress that all the currently spectroscopically selected MBHB candidates have masses of $\gsim 10^8$ M$_\odot$ \citep[e.g.][]{Tsalmantza11,Eracleous12}, and no candidates in the $10^5-10^7$ M$_\odot$ range of interest for the future gravitational wave interferometer LISA have yet been identified. This might be due to a selection effect specific of the traditional spectroscopic search. Indeed, in order to be selected as MBHB candidates, the BELs are typically required to be shifted by $\gsim 1000$ km s$^{-1}$ \citep{Tsalmantza11,Eracleous12} with respect to the host rest frame (traced by the narrow emission lines) in order to prevent the inclusion of single MBHs with slightly asymmetric BLRs. This threshold can be higher than the maximum velocity for which the secondary of a MBHB can retain its own BLR \footnote{As eq.~\ref{eq:mintau}, obtained equating eq.~\ref{eq:bentz2} and eq.~\ref{eq:RL}. For ``solid'' binaries the criterion would be even more constraining.}:  
    \begin{equation}
        v_2 \, \approx \, 480 \, {\rm km \, s}^{-1} \times \left(\frac{M_2}{10^6\,{\rm M}_\odot}\right)^{0.24}\, f_{\rm Edd}^{-0.26} \, f(q)^{-0.5},
    \end{equation}
    where
    \begin{equation}
        f(q) = q^{1/3}(1+q)\left[0.6q^{2/3}+\ln{(1+q^{1/3}})\right],
    \end{equation}
    for small secondary masses and binaries of  close to equal mass, unless the Eddington ratio is small. For example, $M_2=10^6\,{\rm M}_\odot$ and $q=1$ would require $f_{\rm Edd}\lsim 0.01$ to have a maximum secondary velocity of $v_2\approx 1000$ km s$^{-1}$. For such small masses and low $f_{\rm Edd}$, collecting a spectrum with a sufficiently high signal-to-noise ratio to perform the test might be prohibitively challenging. Nevertheless, FUVT  could still succeed in identifying a MBHB (as it does for three mock MBHBs with periods of $ 1000$ yr), if the random fluctuations of the two MBHs do not appear correlated by chance (see the discussion above).
\end{itemize}

As a final note, we stress that the procedure we discuss could, in principle, identify false-positive  MBHB if the structure of the BLR around a single MBH is sufficiently complex, with strong asymmetries, including for example long-lived spiral waves \citep[e.g.][]{storchi03} or hot spots \citep[e.g.][]{fries23}.
If a large and representative distribution of $\overline{CCF}$ were available for bona fide symmetric single MBHs, FUVT could identify subpopulations of outliers. Their deviation from a reference $\overline{CCF}$  distribution would  provide an estimate of the probability that such systems do not belong to the `normal' single MBH distribution (as exemplified in section~\ref{sec:mock}),  allowing the selection of candidates to be further analysed to discriminate between the two (asymmetric single BLR and MBHB) scenarios. We speculate that the inclusion of the information on the cross-correlation of the two (blue and red) parts of the BELs with the observed continuum could inform us about the most plausible scenario. A study quantifying false positives, that is, AGN with single asymmetric BELs, and the development of tests to identify them are ongoing. Nevertheless, we believe that currently the most limiting factor is the paucity of systems for which RM studies have been performed.
Indeed, in order to estimate the probability that our mock MBHBs are outliers of the single MBH distribution, in section~\ref{sec:mock} we have to assume an \emph{a priori}  $\overline{CCF}$ distribution  for single MBHs, as nine measurements are barely enough to characterise the distribution. However, the situation will improve significantly thanks to future large RM campaigns \cite[e.g. the black hole mapper program of the SDSS-V,][]{BHM}, adding approximately $ 1000$ systems to the current sample, allowing far better determination of the distribution of $\overline{CCF}$ for single MBHs. With a large enough sample, one could even begin to identify subpopulations within the single MBH class, ultimately leading to a more accurate identification of binary MBHs candidates. An alternative improvement could be achieved with a dedicated RM campaign on double-peaked emitters\footnote{The two approaches do not necessarily require independent observational campaigns, as double-peaked components might be frequent (or even ubiquitous) in type I AGN \citep{storchi17}.}. Such a study could provide an additional test as to the nature of individual double-peaked emitters \citep[see also][for different tests]{eracleous97, liu16} and, even if all the systems prove to be single MBHs with disk-like BELs \citep[e.g.][]{eracleous94}, could serve as a comparison for future candidates.

\begin{acknowledgements} 
The Authors thank G.~De Rosa and M.~Bentz for their help with the RM datasets, and M.~Eracleous, Z. Haiman and D.J.D' Orazio for their useful comments and suggestions.
MD acknowledge funding from MIUR under the grant PRIN 2017-MB8AEZ, and financial support from ICSC – Centro Nazionale di Ricerca in High Performance Computing, Big Data and Quantum Computing, funded by European Union – NextGenerationEU.
RB acknowledges support through the Italian Space Agency grant \emph{Phase A activity for LISA mission, Agreement n. 2017-29-H.0, CUP F62F17000290005}.

\end{acknowledgements}

%
%

\bibliographystyle{aa} 
\bibliography{main}

\end{document}